\documentclass[journal,draftcls,onecolumn,12pt,twoside]{IEEEtran}
\usepackage{mathrsfs}
\usepackage{amsfonts}
\usepackage{graphicx,cite,epsfig,amssymb,amsmath}
\usepackage{color,xcolor}
\definecolor{red}{rgb}{1.00, 0.00, 0.00}  
\usepackage{pifont}
\usepackage{stmaryrd}
\usepackage{setspace}
\usepackage{subfigure}
\usepackage{cite}
\usepackage{array}
\usepackage{float}
\usepackage{multirow}
\usepackage{algorithm,algpseudocode}
\usepackage{epstopdf}
\usepackage{mathtools}
\usepackage{diagbox}
\usepackage{booktabs}
\usepackage{verbatim}

\providecommand{\algorithmname}{Algorithm}
\floatname{algorithm}{\protect\algorithmname}
\newcommand{\bm}[1]{\mbox{\boldmath{$#1$}}}

\usepackage{arydshln}
\usepackage{enumitem}

\begin{document}
	
\title{Graph Neural Networks Meet Wireless Communications: Motivation, Applications, and Future Directions}
\author{Mengyuan Lee, \IEEEmembership{Graduate Student Member, IEEE,} Guanding Yu, \IEEEmembership{Senior Member, IEEE,} Huaiyu Dai, \IEEEmembership{Fellow, IEEE}, and Geoffrey Ye Li, \IEEEmembership{Fellow, IEEE}
	\thanks{M. Lee and  G. Yu are with the College of Information Science and Electronic Engineering, Zhejiang University, Hangzhou 310027, China. e-mail: \{mengyuan\_lee, yuguanding\}@zju.edu.cn. (\emph{Corresponding author: Guanding Yu})}
	\thanks{H. Dai is with the Department of Electrical and Computer Engineering,
		North Carolina State University, Raleigh, NC 27695 USA. e-mail: hdai@ncsu.edu.}
	\thanks{G. Y. Li is with the School of ECE, Imperial College London, London, UK. e-mail: geoffrey.li@imperial.ac.uk.}}
\maketitle

\begin{abstract}
As an efficient graph analytical tool, graph neural networks (GNNs) have special properties  that are particularly fit for the characteristics and requirements of wireless communications, exhibiting good potential for the advancement of next-generation wireless communications. This article aims to provide a comprehensive overview of the interplay between GNNs and wireless communications, including GNNs for wireless communications (GNN4Com) and wireless communications for GNNs (Com4GNN). In particular, we discuss GNN4Com based on how graphical models are constructed and introduce Com4GNN with corresponding incentives. We also highlight potential research directions to promote future research endeavors for GNNs in wireless communications.
\end{abstract}

\section{Introduction}
With the increasing demand for higher quality of service (QoS) and the explosion of big data, machine learning (ML) based wireless techniques gradually become the mainstream for the sixth-generation (6G) wireless communications and overtake traditional model-based design paradigms \cite{6g}. The reasons are twofold. On the one hand, the mathematical model for wireless communication systems may sometimes fail to accurately model the reality, due to the complex wireless environments such as random channel fading and  interference and the non-linearity induced by the unavoidable hardware impairments. Furthermore, as large-scale wireless communication schemes, e.g. massive multiple-input multiple-output (MIMO), become prevalent, corresponding mathematical models become more complex and related optimization algorithms exhibit higher computational complexity. On the other hand, ML techniques, especially deep learning (DL) techniques, have achieved great success in other fields, such as  natural language processing (NLP) and computer vision (CV), owing to the powerful representation ability and low inference complexity of various neural network models. Therefore, a bunch of recent works have applied DL techniques to wireless communications, such as resource allocation \cite{shi}, physical layer design \cite{physical}, and networking \cite{networking}.

However, there exist several issues when applying DL in wireless communications. First, the neural networks generally require a large number of training samples. However, training samples, either labeled or unlabeled, are usually difficult or expensive to gather in wireless communication systems in general. Meanwhile, too many training samples would make the training process both memory-intensive and time-consuming. Secondly, for most existing works, the structures of neural network models  highly depend on the system size, such as the number of antennas/users. Therefore, these neural network models lack of generalization ability and cannot be used when the system size changes. Finally, most existing ML based wireless algorithms are centralized and may suffer from high signaling cost, possibly severe traffic jam, low scalability and flexibility,  computational limitation, as well as single point of failure.

As an especially proposed neural network model for graphical data,
the graph neural networks (GNNs)  have already achieved good performance in various graph related applications \cite{gnn_survey2} and have the potential to address the above-mentioned challenges.  First, GNNs exploit graph information, especially the graph topological information, more efficiently than other neural network models, which may reduce the number of required training samples. Secondly, GNNs can process input graphs with different sizes. Furthermore, the operations of GNNs are naturally decentralized. As a result, there have been some research works that integrate GNNs with wireless communications recently \cite{graphnn,eisen_graph,shen_graph,unfold,liu_graph,ca,chen,tmc}.

Although there exist some interesting works on GNNs and wireless communications, the interplay between GNNs and wireless communications still needs to be further addressed. Therefore, in this article,  we will provide a clear picture on the interplay between GNNs and wireless communications from two aspects: GNN4Com and Com4GNN. The first aspect, GNN4Com, is about applying GNNs to wireless communications \cite{graphnn,eisen_graph,shen_graph,unfold,liu_graph,ca,chen}.  The second aspect, Com4GNN, is about developing novel wireless mechanisms for the practical implementation of GNNs \cite{tmc}. Although there exist some related overviews \cite{gnn_com_survey1, gnn_com_survey2}, they only focus on GNN4Com and introduce related works in terms of the corresponding applications. This paper pays specific  attention to Com4GNN, including its motivation, necessity, advantage, and existing study. Moreover, the overview on GNN4Com is provided from a new perspective, i.e., based on how graphical models are constructed, including ``wireless networks as graphs" and ``wireless optimization problems as graphs". Besides summarizing and introducing existing works on both GNN4Com and Com4GNN, we also provide  insights and future directions to motivate in-depth investigation.

In what follows, we shall first provide a brief introduction to GNNs in Section II. Next, in Section III, we give an overview of GNN4Com and Com4GNN, respectively. Then, we highlight some potential and important directions for future research in Section IV. Finally, this paper is concluded in Section V.

\section{Graph Neural Networks}
Inspired by the successful applications of convolutional neural networks (CNNs) in various image-based tasks, GNNs are especially designed for graphical data defined over the non-Euclidean domain. By using a bank of convolutional filters, CNNs are able to extract local meaningful features  shared with the entire image datasets, which are powerful for various image analysis. However, the convolutional filters in CNNs only work for image data defined over the Euclidean domain. Therefore, to extract latent representations from graphical data, GNNs are proposed to generalize localized convolutional filters from the Euclidean domain to the non-Euclidean domain. In what follows, we will first introduce how GNNs work for graphical data, and then highlight the advantages and discuss the challenges of implementing GNNs in wireless communications.

\subsection{Basic Operations} 
We use a graph $G(\mathcal{V},\mathcal{E})$ with node set $\mathcal{V}$ and edge set $\mathcal{E}$ as an example to introduce basic principles of GNNs. Depending on specific applications in wireless communications, the nodes in $\mathcal{V}$ can denote access points, base stations, mobile devices, edge servers, and even links between mobile user pairs \cite{graphnn}, while the edges in  $\mathcal{E}$ may indicate whether two nodes can communicate or interfere with each other. Moreover, all nodes and edges may have features, $\bm{x}_v$ and $\bm{x^e}_{(v,u)}$, such as the QoS requirements of mobile devices or the channel gains of interference links.
GNNs take node/edge features and the adjacency matrix of the graph as the input, and output the latent graph/node/edge-level embedding vectors. 

\begin{figure}
\vspace{-2em}
	\centering
	\includegraphics[width=0.9\linewidth, height=0.45\textheight]{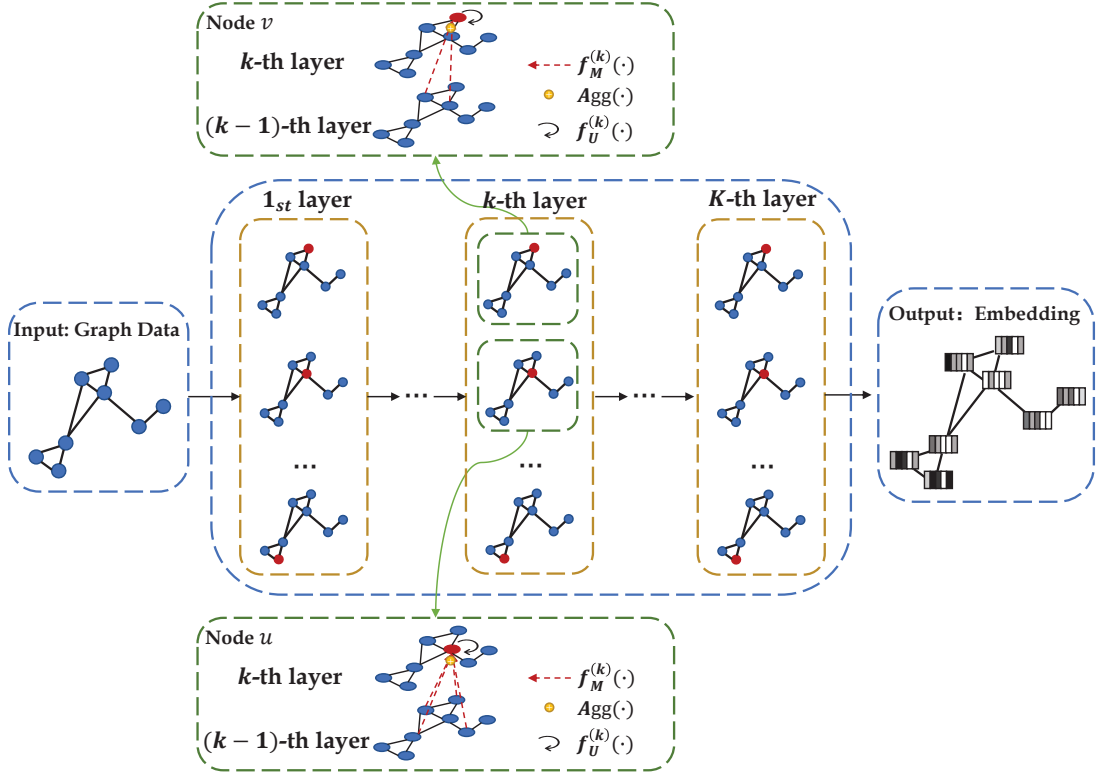}
	\vspace{-1em}
	\caption{A graphical illustration of GNNs' basic operations. Note that the number of neighbors involved in the graph convolutional operation varies with the node degree.}
	\label{fig:gnn}
	\vspace{-2em}
\end{figure}

As shown in Fig. \ref{fig:gnn}, GNNs generally follow the classical layer-wise structure as other neural network models. At the $k$-th layer, the node's embedding vector, $\bm{h}_v^{(k)}$, is updated by aggregating feature information from neighbors $\{u\in N(v)\}$ as follows
$$\bm{h}_v^{(k)}= f_U^{(k)}(\bm{h}_v^{(k-1)}, {\rm Agg}_{u\in N(v)}(f_M^{(k)}(\bm{x^e}_{(v,u)},\bm{h}_v^{(k-1)},\bm{h}_u^{(k-1)}))), $$
where  $\bm{h}_v^{(0)}=\bm{x}_v$, $f_M^{(k)}(\cdot)$ and $f_U^{(k)}(\cdot)$ are respectively local message and update functions parameterized by neural networks, which are defined at the node-level and whose parameters are shared by all nodes in the graph. ${\rm Agg}(\cdot)$ is the aggregation function whose popular candidates include $\sum(\cdot)$, ${\rm mean}(\cdot)$, and $\max(\cdot)$. The graph convolutional operation in GNNs is similar to that in CNNs, where the parameters of convolutional filters are also trainable. However, the aggregation region in CNNs is a fixed-size square area (such as 3$\times$3) while that in GNNs dynamically varies with the node degree as suggested by Fig. \ref{fig:gnn}. This is achieved by the fact that  ${\rm Agg}(\cdot)$ can take varying numbers of vectors as inputs and the specific number of input vectors depends on the node degree. Therefore, GNNs can operate on graphical data defined over non-Euclidean domains. 

With the help of the graph convolutional operation, at each layer of GNNs,  each node collects information from its one-hop neighbors. By stacking multiple layers, the operation repeats and information propagates through the network. The final embedding vectors can be input into other machine learning techniques (e.g. multi-layer perceptions \cite{graphnn} and reinforcement learning \cite{chen}) for further node-level analysis, such as a device's on/off prediction. Furthermore, with the help of an additional readout function, all nodes' embedding vectors can be summarized into a graph-level embedding vector for graph-level tasks, such as the network's throughput prediction.

\subsection{Advantages of GNNs for Wireless Communications}
GNNs are efficient graph analytical tools. They show satisfactory performance in various applications, such as CV, NLP, traffic management, recommendation systems, and protein analysis. By constructing graphical models for wireless networks, GNNs can be naturally applied to wireless networks. More importantly, GNNs have several special properties that are particularly fit for the characteristics and requirements of wireless communications. They are of great potential to overcome the  three aforementioned challenges about employing ML in wireless communications.

First, GNNs learn low-dimensional vector representation by extracting node/edge features and topological information simultaneously. However, topological information is of high dimension, and thus difficult to be fully exploited by traditional mathematical  techniques and other neural network models, such as CNNs and recurrent neural networks (RNNs). Therefore,  using GNNs can potentially achieve better performance with fewer training samples for various tasks in wireless communications.

Secondly, the parameters of each GNN layer are shared among all nodes in the graph, i.e., $f_M^{(k)}(\cdot)$ and $f_U^{(k)}(\cdot)$ are invariant for different nodes.  Therefore, GNNs have good generalization ability concerning input size, which is particularly fit for  the dynamic nature of wireless communications.

Finally, GNNs update the embedding vectors of nodes based on neighborhood information only. Therefore, once  a GNN model is trained and deployed, the inference stage can be  implemented in a decentralized manner. To be more specific, each node only needs to exchange information with its neighbors and can get its own prediction result locally.  Note that most neural network models widely adopted in wireless communications, such as CNNs and RNNs, require central processing during the inference stage. Therefore, GNNs can facilitate decentralized control and resource management, which is appealing for large-scale wireless communication systems.

\subsection{Challenges of GNNs for Wireless Communications}
Although GNNs are a natural and powerful tool for wireless communications, there exist several challenges. The first is constructing appropriate graphical models for wireless networks, which is the premise of applying GNNs in wireless communications (i.e. GNN4Com) and will be further discussed in Section III.A. The second is about implementing the inference stage of GNNs in a decentralized manner for wireless networks to achieve decentralized control and resource management. As introduced above, GNNs update each node's embedding by aggregating neighborhood information. Therefore, the information exchange among neighbors is inevitable and generally realized through imperfect wireless communications. Moreover, aggregating neighborhood information makes all nodes in the graph  entangled, which not only makes GNNs more vulnerable to noise and errors than other neural networks but also entails stringent   synchronization requirements. This imposes great burden on wireless transmission mechanism design and indicates the necessity of Com4GNN, which will be further discussed in Section III.B.

\section{Graph Neural Networks and Wireless Communications}
\begin{figure}
	\vspace{-2em}
	\centering
	\includegraphics[width=0.7\linewidth, height=0.25\textheight]{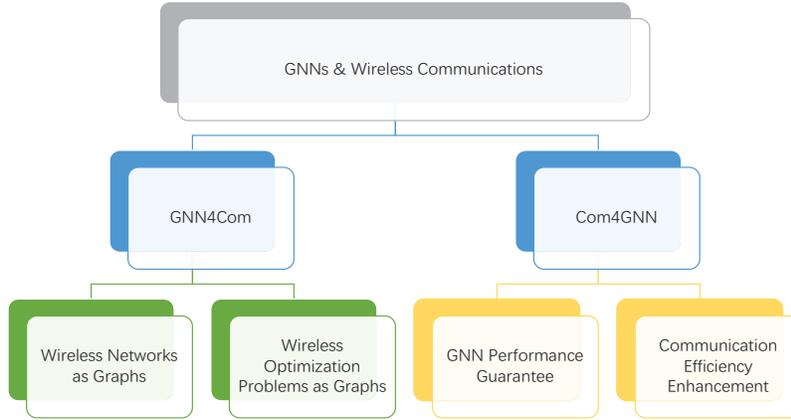}
	\vspace{-1em}
	\caption{Interplay between GNNs and wireless communications.}
	\label{fig:structure}
	\vspace{-1em}
\end{figure}

In this section, we give an overview about the interplay between GNNs and wireless communications, consisting of two aspects: GNN4Com and Com4GNN as shown in Fig. \ref{fig:structure}. On the one hand, as a powerful tool for wireless communications, GNNs have been applied to solve various tasks in wireless communications, including link scheduling \cite{graphnn}, power control \cite{eisen_graph,shen_graph,unfold}, user association \cite{liu_graph}, and other emerging fields \cite{ca,chen}. On the other hand, to guarantee the inference accuracy, robustness, and latency when deploying GNNs for decentralized inference in wireless communications, novel wireless mechanisms and protocols \cite{tmc} are indispensable as mentioned above. 

\subsection{Graph Neural Networks for Wireless Communications}
To give a picture on GNN4Com from a new perspective, we categorize the existing works based on how graphical models are constructed, corresponding to the first challenge in Section II.C.

\subsubsection{Wireless Networks as Graphs}
The most natural way is to construct directed/undirected graphs based on the structure of the considered wireless networks by regarding devices as nodes and channels as edges. After that, GNNs are utilized in an end-to-end paradigm to learn the relationships between graphical models and output variables \cite{graphnn,eisen_graph,shen_graph,liu_graph,chen}. We use the GNN-based link scheduling algorithm in \cite{graphnn} as an example to illustrate the above paradigm.

Link scheduling for device-to-device (D2D) communications is usually formulated as an NP-hard non-convex combinatorial problem, aiming at maximizing the network utility by activating a subset of mutually interfering links, i.e., D2D pairs. Traditional methods are based on various mathematical optimization techniques by estimating accurate channel state information (CSI), incurring huge overhead both in time and resource. Therefore, the link scheduling algorithm independent of CSI is desired, where the link scheduling problem cannot be formulated mathematically. Fortunately, data-driven DL techniques are possible solutions. Moreover, topology and distance information to some extent can remedy the lack of CSI. Therefore, GNNs\footnote{In the original manuscript of \cite{graphnn}, the adopted technique is termed as ``graph embedding". According to \cite{gnn_survey2}, deep learning based graph embedding unifies graph embedding and GNNs. Therefore, in this paper, we term the technique adopted in \cite{graphnn} as GNNs.} have been effectively used in \cite{graphnn} to deal with the link scheduling problem.

\begin{figure}
	\vspace{-2em}
	\centering
	\includegraphics[width=0.8\linewidth, height=0.25\textheight]{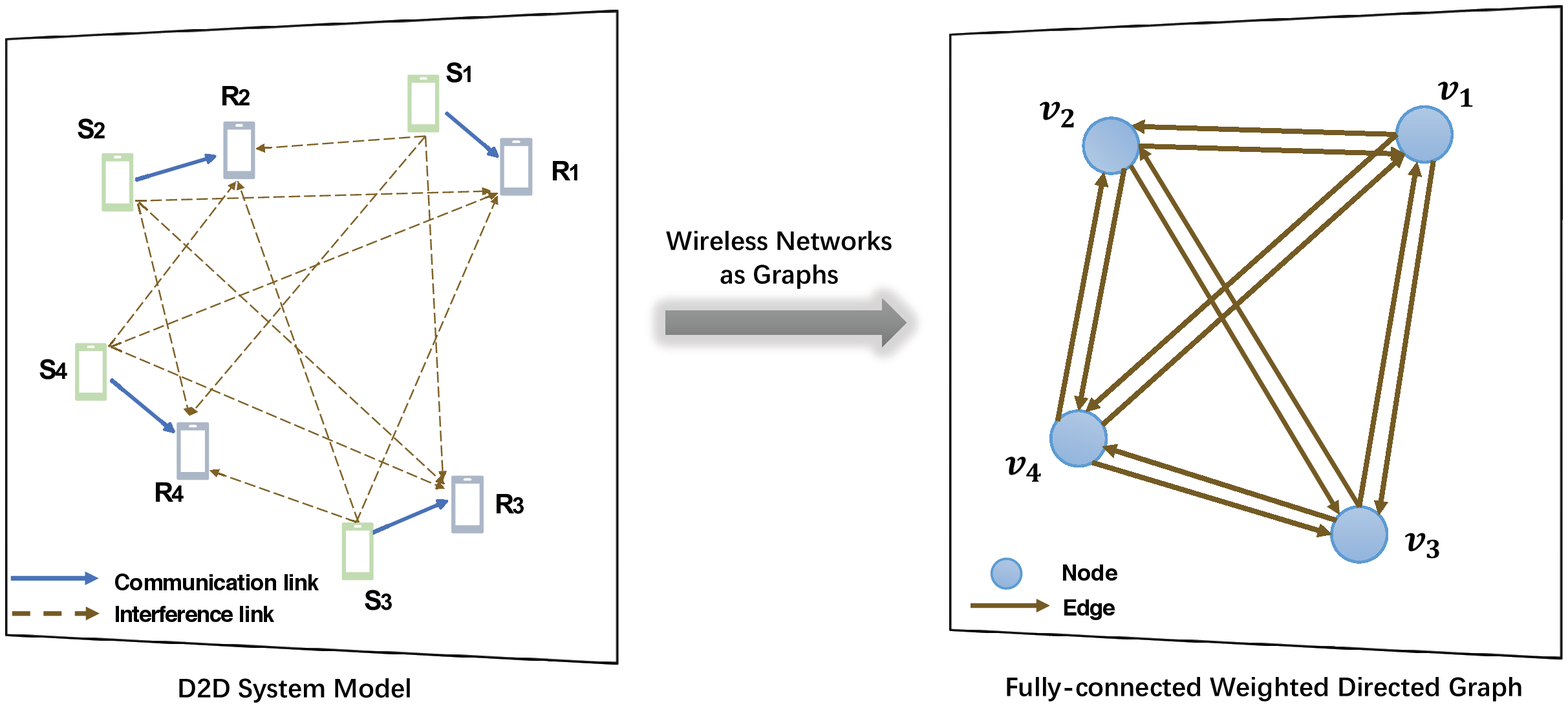}
	\vspace{-1em}
	\caption{Wireless networks as graphs: Graphical model for D2D communication systems in \cite{graphnn}.}
	\label{fig:d2d}
	\vspace{-2em}
\end{figure}

By following the ``wireless networks as graphs" paradigm, each D2D pair is regarded as a node \cite{graphnn}. Two D2D pairs are connected by an edge if they interfere with each other. Note that each edge has a direction, which corresponds to the direction of interference. Moreover, the node and edge features depend on the distances between the two devices of the corresponding communication/interference links. In this way, the original wireless D2D network can be represented as a fully-connected weighted directed graph as depicted in Fig. \ref{fig:d2d}. By taking the graphical model in Fig. \ref{fig:d2d} as the input, GNNs are utilized to learn the vector representation of each D2D pair based on topology and distance information. Since the link scheduling indicator is binary, the output embedding vectors of each D2D pair are then input to a fully-connected neural network for binary classification. By performing simulation on a 500 m by 500 m two-dimensional square area with 50 D2D pairs, the GNN-based link scheduling algorithm converges after only 30-40 training epochs with 500 training network layouts, and the systems' sum rate only suffers from 3.15\% loss compared to the state-of-the-art algorithm that needs 800,000 training network layouts. Additional simulation results on scenarios with different topologies validate its scalability and generalization ability, which are preferred properties in wireless communications.

Besides following the end-to-end learning paradigm, ``wireless networks as graphs" can also be used in the model-driven paradigm. For example, an unfolded version of the weighted minimum mean square error (WMMSE) algorithm for power allocation in wireless communications has been proposed in \cite{unfold}, where the learnable modules are parameterized via GNNs. The GNN-based unfolded WMMSE algorithm achieves good performance and generalization ability in networks with different densities and sizes.

\subsubsection{Wireless Optimization Problems as Graphs}
It generally works well to construct graphical models based on the structure of wireless networks. However, for ultra-dense and heterogeneous networks, the graphical models based on network structures are usually not only heterogeneous but also of large-size, which imposes a huge burden on the prediction ability of GNNs. Alternatively, the graphical model can be constructed based on the mathematical formulation of a specific optimization problem in the considered wireless networks. 

\begin{figure}
\vspace{-2em}
	\centering
	\subfigure[Graphical model.]{
		\begin{minipage}[t]{0.6\linewidth}
			\centering
			\includegraphics[width=0.95\linewidth]{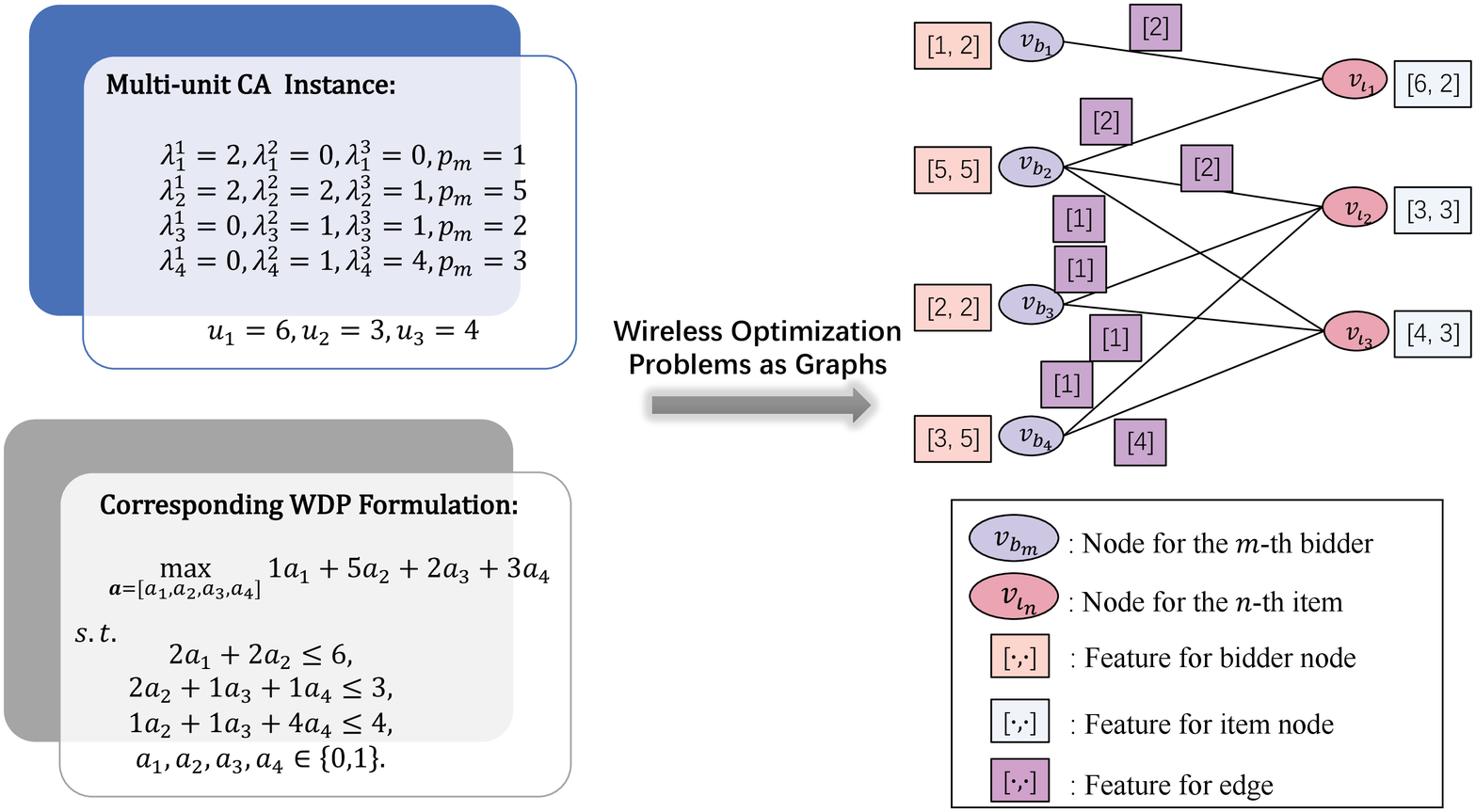}
			\label{fig:tcc}
			\vspace{-1em}
		\end{minipage}
	}
	\subfigure[Simulation results.]{
		\begin{minipage}[t]{0.35\linewidth}
			\centering
			\includegraphics[width=0.95\linewidth]{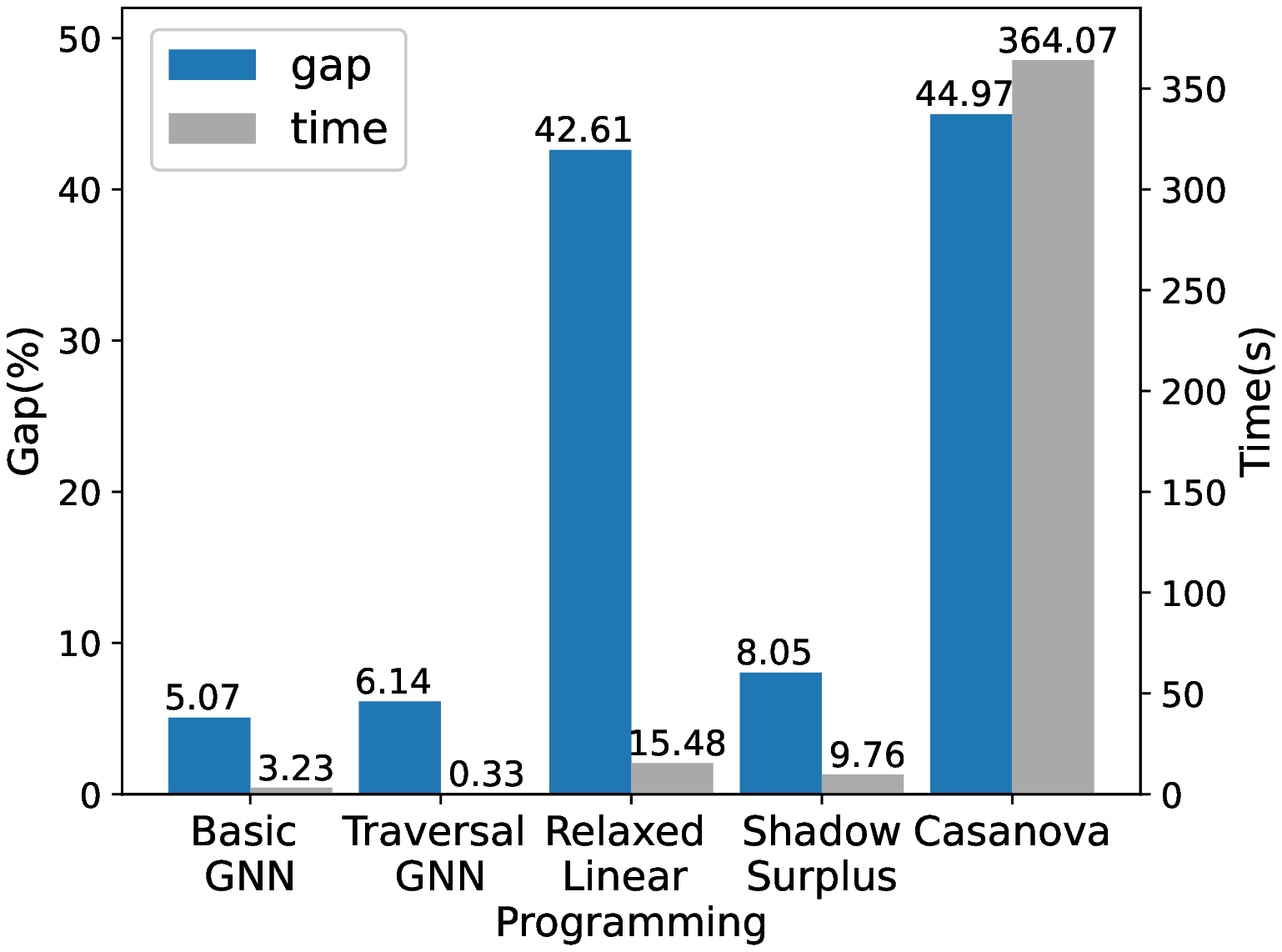}
			\label{fig:tcc_simu}
			\vspace{-1em}
		\end{minipage}
	}
\vspace{-1em}
	\caption{Wireless optimization problems as graphs: Graphical model and simulation results for the WDP in the multi-unit CA in \cite{ca}.}
	\label{fig_tcc}
	\vspace{-2em}
\end{figure}

We adopt \cite{ca} as an example to illustrate this alternative paradigm of using GNNs in wireless communications. This work addresses the multi-unit combinatorial auction (CA), an efficient  resource allocation mechanism in various fields including wireless communications. It is about how to allocate $N$ items (e.g. bandwidth, time slot), $\{\iota_1, \iota_2, ..., \iota_N\}$, among $M$ bidders (e.g. interfering users), $\{b_1, b_2, ..., b_M\}$, where each item $\iota_n$ has limited stock denoted as $u_n$. Specifically, each bidder $b_m$ first proposes its bid to the auctioneer (e.g. base station and edge server), where the bid comprises the quantity that bidder $b_m$ requests for each item, $ \{\lambda_m^1, \lambda_m^2, ..., \lambda_m^N\}$, and the maximum price/contribution (e.g. throughput), $p_m$, that bidder $b_m$ pays for/achieves with the bundle. After that, with the principle that the bundle requested by each bidder cannot be partially satisfied, the auctioneer needs to make allocation decision and output a binary allocation vector, $\bm{a}=[a_1, a_2, ..., a_M] \in [0,1]^M$, to maximize the overall performance (e.g. weighted sum rate) denoted as $\sum_{m=1}^M p_m a_m$.  This allocation problem is termed as the winner determination problem (WDP), which is NP-complete and difficult to solve. An instance of the WDP in multi-unit CA can be found in Fig. \ref{fig:tcc}.

The work in \cite{ca} proposes an efficient and scalable GNN-based method for the WDP, where the interactions between the bidders and items are important information and need to be included in the graphical model. However, if we regard all bidders and auctioneer as nodes by following the ``wireless networks as graphs" paradigm, the information about items is difficult to be included in the graphical model and the interactions between bidders and items cannot be reflected. Alternatively, by investigating  the mathematical formulation of the WDP, an augmented bipartite bidder-item graphical model is proposed \cite{ca}. As illustrated in Fig. \ref{fig:tcc}, the proposed graphical model consists of two sets of nodes, i.e., $\mathcal{V}_b$ and $\mathcal{V}_\iota$, corresponding to the two index sets, $\{1,2,...M\}$ and  $\{1,2,...N\}$. To be more specific, each node in $\mathcal{V}_b$ represents a bidder while that  in $\mathcal{V}_\iota$ represents an item. Moreover, edges only exist between nodes from different sets. Specifically, if bidder $b_m$ requests item $\iota_n$, the bidder node $v_{b_m}$ and the item node $v_{\iota_n}$ are connected by an edge. Furthermore, node and edge features are designed to incorporate all parameters in the mathematical formulation of the WDP in the graphical model.

After constructing graphical  models, GNNs are utilized to learn the probability of each bidder node included in the optimal allocation. Since the proposed augmented bipartite bidder-item graphical model is heterogeneous, half graph convolutional operation is utilized,  where  the $f_M^{(k)}$ and $f_U^{(k)}$ mentioned in Section II.A operate on item and bidder nodes alternately.  At the output layer, only the embedding vectors of bidder nodes are input to a softmax layer for  probability  prediction. Different from the operations introduced in Section II.A that focus on the outputs of all nodes, the structure of  GNNs with half graph convolutional operations avoids unnecessary computational complexity. With the help of basic and traversal post-processing algorithms, the continuous probability output by GNNs is transformed into WDP's binary allocation result. Through extensive simulation on synthetic instances where $M$ ranges from 1,000 to 5,000 and $M/N=10$ always holds, two metrics, i.e., the performance gap and the execution  time, are adopted, where performance gap refers to the revenue loss compared to the optimal solution. The simulation results in Fig. \ref{fig:tcc_simu} indicate that the GNN-based method can achieve a smaller performance gap with less execution  time compared with the state-of-the-art heuristic algorithms, and also has good generalization ability concerning problem size.

\subsubsection{Summary and Insights}
The ``wireless networks as graphs" paradigm is the mainstream of using GNNs in wireless communications. When one wants to apply GNNs to a new wireless application, one can first try this graphical model construction paradigm combined with \textbf{carefully selected or designed  features}. If the performance is not satisfactory, one can then turn to the ``wireless optimization problems as graphs" paradigm, which generally consists of two steps: {\bf{defining node sets}} and {\bf{including parameters as features}}. For some applications, such as the WDP, the latter graph modeling paradigm has more powerful representation ability than the former one.  We believe that,  with the inclusion of this alternative paradigm, GNNs will have wider applications in large-scale heterogeneous wireless networks.

\subsection{Wireless Communications for Graph Neural Networks}
To deal with the second challenge in Section II.C, developing novel wireless mechanisms is crucial, which indicates the necessity of studying Com4GNN. Specifically, Com4GNN is about exploiting the characteristics of GNN inference tasks and designing task-oriented wireless transmission mechanisms. Compared with existing wireless mechanisms that aim at reliable transmission, Com4GNN has the potential to not only enhance the performance of GNNs but also decrease the communication overhead. However, even if it is an important topic, the study on Com4GNN is relatively rare compared with that on GNN4Com. In the following, we will introduce an existing work that develops a novel retransmission mechanism for GNNs. Based on this, we then suggest  two incentives  to promote future study on Com4GNN.

As mentioned above, to deploy GNNs for decentralized inference in wireless communications, the information exchange among neighbors is generally realized through noisy and fading wireless channels. The imperfect wireless transmission would lead to received signals with random errors, and eventually deteriorate the inference accuracy of GNNs. This is the main bottleneck for achieving decentralized control and management in wireless communications with GNNs. In \cite{tmc}, a novel retransmission mechanism is proposed to deal with this issue. 

To guarantee the accuracy and robustness of the predicted results, a robustness verification methodology is developed prior to designing retransmission mechanisms. For simplicity, a decentralized GNN binary classifier serves as an example and a robustness verification problem is formulated, whose solution indicates whether the predicted label is robust with a certain transmission error bound, i.e., the maximum number of tolerable errors in each received signal from the neighbors.
\begin{figure}
\vspace{-2em}
	\centering
	\subfigure[Basic idea.]{
		\begin{minipage}[t]{0.6\linewidth}
			\centering
			\includegraphics[width=0.95\linewidth]{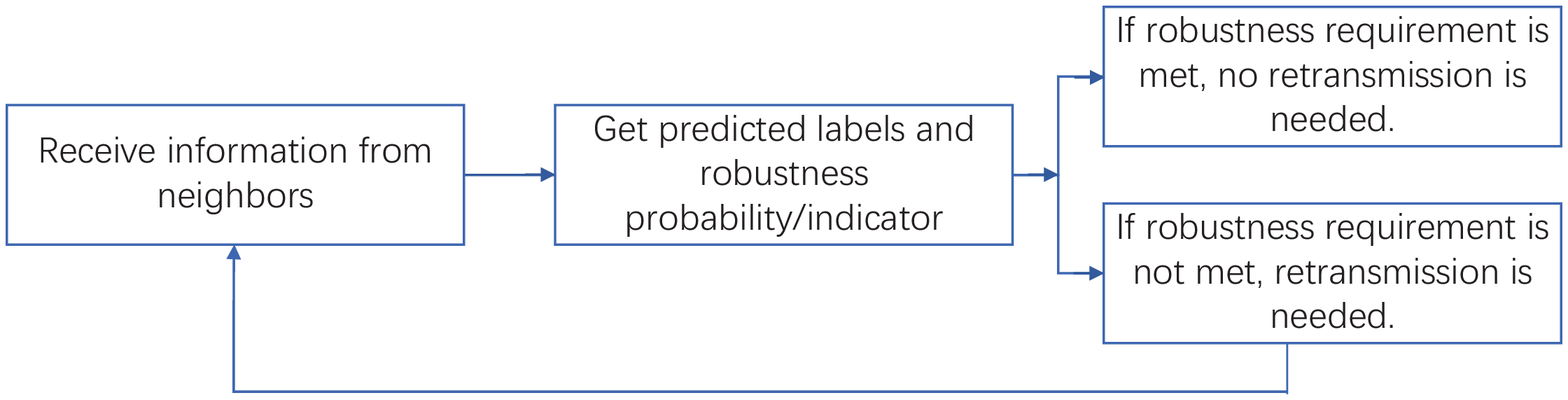}
			\label{fig:flow}
			\vspace{-1em}
		\end{minipage}
	}
	\subfigure[Simulation results.]{
		\begin{minipage}[t]{0.35\linewidth}
			\centering
			\includegraphics[width=0.95\linewidth]{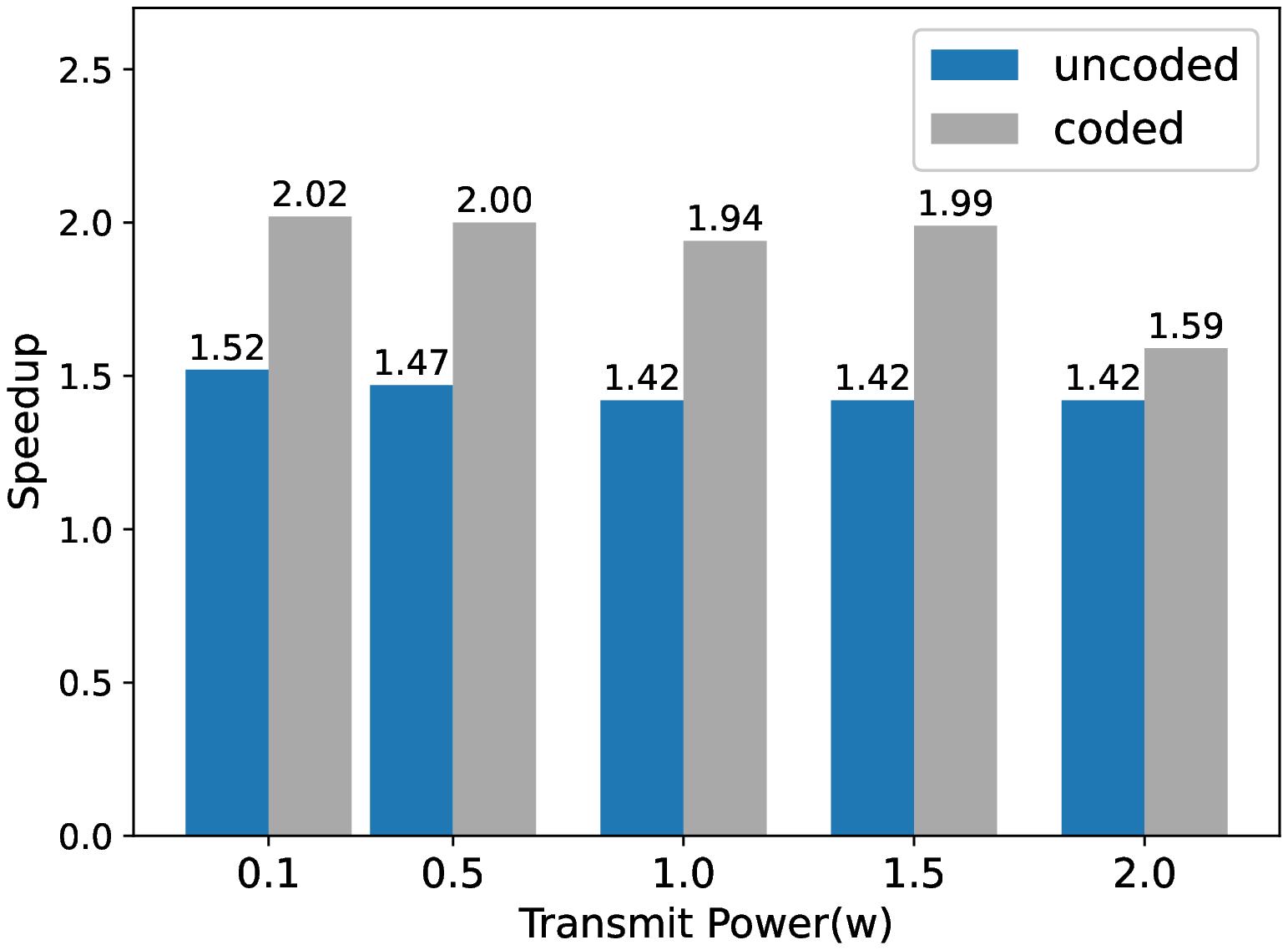}
			\label{fig:gain}
			\vspace{-1em}
		\end{minipage}
	}
\vspace{-1em}
	\caption{Basic idea and simulation results of the proposed retransmission mechanism in \cite{tmc} for robust and accurate decentralized GNN inference.}
	\label{fig_tmc}
\vspace{-2em}
\end{figure}
In uncoded wireless communication systems, the transmission error bound follows the binomial distribution parameterized by the bit-error rate (BER). Based on this fact, the robustness probability of the prediction result can be derived, which is a function of BER and serves as the retransmission criterion. To be more specific, retransmission is needed when the robustness probability of the predicted label falls below the pre-given target robustness probability. In coded wireless communication systems, the transmission error bound equals either 0 or the length of the codeword, depending on whether the signal transmission is in outage. Following this analysis, a binary robustness indicator of the prediction result is proposed and serves as the  retransmission criterion. Specifically, retransmission is needed when the prediction robustness indicator is 0. The above process is summarized as the flowchart in Fig. \ref{fig:flow}. 

Note that traditional retransmission mechanisms aim at reliable transmission, and  retransmission is needed whenever the BER threshold is not met or at least one neighbor is still in outage for uncoded and coded wireless communication systems, respectively. However, it neglects that GNNs have some tolerance to the errors in received signals from neighbors and inevitably incurs overwhelming communication overhead. In fact, retransmission does not need to be invoked each time and can stop much earlier in many cases. With the novel task-oriented retransmission criterion proposed in \cite{tmc}, the error tolerance of GNNs is fully exploited, and thus fewer transmission rounds are needed to achieve robust and accurate prediction with decentralized GNNs. To validate the above advantage, simulation is conducted on the scenarios where the decentralized GNN binary classifier is implemented on 200 users with Rayleigh fading channels. The simulation results are summarized in Fig. \ref{fig:gain}, where the ratio between the average number of needed retransmission rounds of the traditional and the proposed retransmission mechanisms, i.e., speedup, is adopted as a metric. The simulation results indicate that  neglecting GNNs' inherent robustness and directly adopting existing wireless transmission mechanisms/protocols will lead to unnecessary overhead.

We summarize this part by again emphasizing the potential benefits and incentives of studying Com4GNN.
\begin{itemize}
	\item {\bf{GNN performance guarantee}}: The decentralized prediction performance of well-trained GNN models is hindered by imperfect wireless transmission, and wireless transmission mechanisms/protocols can be carefully designed  to address this issue. Therefore, studying Com4GNN will help guarantee the performance of GNNs.
	\item {\bf{Communication efficiency enhancement}}: GNNs themselves have some robustness to errors and noises, and other structural characteristics, which is neglected by traditional wireless design. Therefore, the communication efficiency may be further enhanced through task-oriented Com4GNN.
\end{itemize} 
Overall, Com4GNN is worth in-depth investigation.

\section{Future Directions}
Both GNN4Com and Com4GNN have attracted increasing attention from the research community in recent years. We believe that the following research directions are worth exploring and are crucial to bring GNN-based wireless techniques from theory to practice.
\subsection{New Wireless Applications of GNNs}
Existing works on  GNN4Com generally focus on resource allocation in wireless communications. The potential of GNNs in other wireless applications, especially the physical layer communications, has not been fully exploited. Using CSI feedback in the massive MIMO system as an example, existing methods are generally developed with the help of CNNs or forward neural networks (FNNs), which cannot be generalized to systems with dynamic number of antennas. However, the operations of GNNs are independent of input size, making it a good candidate for developing antenna number-independent CSI feedback methods.

On the other hand, existing works on GNN4Com usually focus on a certain network layer of wireless communications and ignore the relationship between different network layers. However, joint design over different network layers is important to achieve  performance gain of the whole communication systems. Consequently, how to incorporate GNNs into cross-layer optimization is worth future investigation. 

\subsection{Efficient GNN Training Strategy}
Generally speaking, the training stage of neural networks, including GNNs, are both time-consuming and resource-intensive.  Most existing works on GNN4Com implement the training stage of GNNs in the offline mode. However, the overhead of GNN training, even if offline, is inevitable and may become the bottleneck of applying GNN4Com techniques in practice. 

Possible solutions include the following aspects. First, collaborative learning schemes, such as edge learning and federated learning, can be used to train GNN models by exploiting the computation resources of multiple devices. Secondly, it is also appealing to further reduce the number of required training samples of GNNs with data augmentation and contrastive learning techniques. Furthermore, online GNN training, GNN model pruning, or GNN model improvement are worth in-depth study. In summary, enhancing the training efficiency of GNNs in wireless networks is another important future direction.

\subsection{Wireless Mechanism/Protocol for GNN Inference}
As mentioned above, the inference accuracy of GNNs in a decentralized manner is  inevitably deteriorated by the fading and noisy wireless channel. Developing wireless mechanisms/protocols for robust decentralized GNN inference is a key issue to bring GNNs to practice. In \cite{tmc}, a retransmission mechanism is proposed to achieve the above goal. One can also consider other techniques, such as power control, adaptive modulation and coding,  joint source-channel coding, and error correction mechanisms to further facilitate decentralized GNN inference in wireless communications. 

In addition to remedy imperfect wireless transmission, some wireless mechanisms/protocols are prompted by the special structures of GNNs. For example, for multi-layer GNNs, embedding vectors at different hidden layers are supposed to be shared among neighbors through wireless transmission. Therefore, the synchronization and transmission delay among neighbors need to be controlled. How to develop effective synchronization protocols and allocate resource  among neighbors is also important to facilitate the practical implementation of GNNs in wireless communications.

\subsection{Privacy-Preserving GNN Training/Inference}
Privacy preserving is one of the core issues for future wireless communications. However, both the training and inference procedures of GNNs may lead to privacy leakage.
 
For the training process of GNNs, training samples are indispensable, which are generally obtained from system log files or users' local data. This process inevitably exposes parts of users' private information. A possible solution is to use a noisy version of the original data (e.g., through differential privacy), and how to train GNNs with noisy samples needs further investigation. Moreover, the above-mentioned collaborative learning scheme, such as edge learning and federated learning, may help protect users' privacy during the training process.
 
As for the inference stage, the decentralized GNN inference relies on information exchange among neighbors, which also leads to the privacy issue. To overcome this challenge, possible solutions include developing  privacy-preserving communication techniques and modifying the operations of GNNs. In brief, achieving decentralized control with GNNs while protecting user privacy is challenging but desired in practice.

\section{Conclusions}
This article provides a whole picture of the interplay between GNNs and wireless communications, including GNN4Com and Com4GNN. Compared to the existing works, we are among the first to introduce the concept and applications  of Com4GNN. Moreover, for both aspects, we provide an overview on the existing works from new perspectives. Some important insights and potential directions are also highlighted. Overall speaking, this article is among the first attempts to offer useful guidance and insightful directions for future research endeavors concerning the application of GNNs in wireless communications. It is expected that  GNNs will play an increasingly important role in the next generation wireless communication systems.

\end{document}